# Formation of Heliospheric Arcs of Slow Solar Wind

A. K. Higginson[1]*, S. K. Antiochos[2], C. R. DeVore[2], P. F. Wyper[3], & T. H. Zurbuchen[1]**

[1]Department of Climate and Space Sciences and Engineering, University of Michigan, Ann Arbor, MI 48109, USA.

[2]Heliophysics Science Division, NASA Goddard Space Flight Center, Greenbelt, MD 20771, USA.

[3]Department of Mathematical Sciences, Durham University, Durham DH1 3LE, UK.

*Correspondence to: aleida@umich.edu

** Now at NASA Headquarters, Washington, DC 20546, USA.

# ABSTRACT

A major challenge in solar and heliospheric physics is understanding the origin and nature of the so-called slow solar wind. The Sun's atmosphere is divided into magnetically open regions, coronal holes, where the plasma streams out freely and fills the solar system, and closed regions, where the plasma is confined to coronal loops. The boundary between these regions extends outward as the heliospheric current sheet (HCS). Measurements of plasma composition strongly imply that much of the slow wind consists of plasma from the closed corona that escapes onto open field lines, presumably by field-line opening or by interchange reconnection. Both of these processes are expected to release closed-field plasma into the solar wind within and immediately adjacent to the HCS. Mysteriously, however, slow wind with closed-field plasma composition is often observed in-situ far from the HCS. We use high-resolution, three-dimensional, magnetohydrodynamic simulations to calculate the dynamics of a coronal hole whose geometry includes a narrow corridor flanked by closed field and which is driven by supergranule-like flows at the coronal-hole boundary. These dynamics produce giant arcs of closed-field plasma that originate at the open-closed boundary in the corona, but extend far from the HCS and span tens of degrees in latitude and longitude at Earth. We conclude that such structures can account for the long-puzzling slow-wind observations.

# 1. Introduction

## 1.1 Slow Solar Wind

The actions of the Sun's magnetic field create both the hot (1–2 MK) solar corona and the supersonic solar wind (Parker 1958). In regions where field lines start and end at the photosphere, plasma is confined to form the bright extreme ultraviolet (EUV) loop structures of the "closed" corona (Orall 1981). In "open" regions, field lines start on the Sun but extend out into the solar system. These regions, which appear dark in EUV and are referred to as "coronal holes" (Zirker 1977), are the sources of the fast wind, which has speeds > 500 km s$^{-1}$ (Zurbuchen 2007). The slow wind, in contrast, has speeds < 500 km s$^{-1}$ (Schwenn 1990). Compositional studies indicate that the slow wind originates from regions with temperature and elemental abundances that much more closely match the properties of the hot, closed corona than the cooler, open coronal holes (Zurbuchen 2007).

Observations strongly imply that slow-wind plasma originates in the closed-field regions, but then escapes onto open field lines and propagates into the heliosphere. Given that cross-field diffusion is expected to be negligible, the only processes by which closed-field plasma can escape require magnetic-field dynamics: either rapid opening of closed field, or magnetic reconnection of open and closed flux (interchange reconnection; Crooker et al. 2002). Because both processes are intrinsically dynamic, they imply that the slow wind should exhibit non-steady dynamics and, indeed, recent observations have confirmed that the slow wind has a distinctly dynamic structure (Kepko et al. 2016).

The location of the slow wind also suggests that it is associated with closed-field regions. Both processes above occur only at the boundary between open and closed flux in the corona. This boundary extends outwards as the heliospheric current sheet (HCS) (Smith 2001). If the source of the slow wind is the open-closed boundary, then this wind should be located near the HCS and, in fact, observations show that the HCS is always imbedded in slow wind, never fast (Burlaga et al. 2002).

The longstanding puzzle, however, is that the slow wind is frequently found far from the HCS. This surprising trend was first observed and is well documented by in-situ observations (Zurbuchen 2007). This poses a major conundrum: How can slow wind originate at or very near the HCS at the Sun, but be found very far from the HCS in the heliosphere?

**1.2 Separatrix-Web Theory**

The Separatrix Web (S-Web) theory (Antiochos et al. 2011) offers a resolution to this paradox. Underlying the theory is the fact that coronal holes frequently have the specific geometry in which extended polar and low-latitude open regions are connected by a narrow corridor, as shown in the *Solar Dynamics Observatory* EUV images in Figure 1A,B. In this geometry, the mapping of open flux into the heliosphere is quasi-singular: some open field lines very near the open-closed boundary map to locations far from the HCS. A simple representation of this is shown in Figures 1C and 2, where several such field lines are represented by cyan dots on the surface (1C) and corresponding cyan lines in the inner heliosphere (2). The black field lines in Figure 1C outline the coronal-hole boundary, indicating the presence of the corridor connecting the polar and low-latitude open regions. This "elephant trunk" invariably forms whenever low-latitude active regions appear at the solar surface (Zirker 1977).

We used a numerical magnetohydrodynamic (MHD) model, described below, to calculate the quasi-steady solution for the corona and wind shown in Figure 2. This solution shows explicitly how the magnetic field maps from the open-field corridor into the heliosphere. In the figure, the yellow sphere is the solar surface at $1R_\odot$. The current density (gray-scale contours) in the ecliptic plane highlights the location of the HCS. Figure 1C shows a line (cyan dots) drawn on the surface across the narrowest part of the corridor. The cyan field lines in Figure 2 are traced from this line. Although the line of dots in Figure 1C is very short on the surface, its magnetic mapping into the heliosphere forms a giant arc in Figure 2 spanning tens of degrees in latitude and longitude. The dots in the corridor's center map to the highest-latitude field lines in the arc; those at the ends of the line map to very near the HCS. Flux from the polar open region maps to everywhere above the cyan arc, whereas flux from the low-latitude region maps to the volume between the arc and the HCS. As long as the ratio of fluxes in the low-latitude and polar open regions remains relatively constant, the arc in the heliosphere maintains its large angular extent, *irrespective of the narrowness of the corridor*. This result is the primary motivation for the S-Web theory of the high-latitude slow wind: closed-field plasma can be released all along this arc due to dynamics localized at the open-closed boundary.

In a steady-state model with the configuration in Figure 2, all closed field lines remain closed, and there is no slow-wind plasma anywhere. The coronal magnetic field, however, is constantly being stirred by convective motions, in particular by the large cells known as supergranules, which are ever-present on the surface. Supergranules have a size scale of 30,000 km, lifetimes around 30 hr, and typical horizontal velocities of 1 km s$^{-1}$ (Bray & Loughhead 1967). The S-Web model conjectures that the supergranular driving of the magnetic field at an open-field corridor, illustrated in Figure 1, results in dynamics at the open-closed boundary that release closed-field plasma all along the heliospheric arc of Figure 2. As described by Antiochos et al. (2011), the heliospheric-field configuration of coronal holes is routinely far more complex than Figure 2. Rather than creating a simple HCS with one separatrix arc, the solar surface flux distribution creates an intricate web of separatrices ("S-Web"). Therefore, slow wind should be released along not just a single high-latitude arc, but across a broad network of such arcs. If correct, this would naturally account for the observation of slow wind far from the HCS; however, it seems unlikely that the conjecture is valid. The critical magnetic-field dynamics are the processes of closed flux being opened and interchange reconnection, described above, and these must occur at the open-closed boundary; that is, where the arc intersects the HCS. How can such dynamics release closed-field plasma at the arc's center, which is tens of degrees in angular separation from the HCS? To answer this question, we performed the first numerical simulations of an open-field corridor driven by photospheric motions, to determine whether the resulting dynamics can resolve the longstanding slow-wind conundrum.

## 2. Numerical Model

For our simulations, we use the Adaptively Refined Magnetohydrodynamics Solver (ARMS) as described by Higginson et al. (2017), using the initial magnetic field distribution of Antiochos et al. (2011). To simulate the supergranular driving of this field, we apply a localized rotational velocity (Figure 1C) on the surface at $1R_\odot$ (as in Higginson et al. 2017) that spans the corridor.

The sub-Alfvénic flow ($v_{max} < 9$ km s$^{-1}$) is turned on just long enough to rotate field lines through 180°, from one side of the corridor to the other. Real supergranular motions are far more complex than a simple half turn, but if our idealized driving can account for the slow wind, then the quasi-chaotic driving of the Sun's supergranular flows must be even more effective.

## 3. Results

The photospheric driving has three main effects on the coronal magnetic field. In open regions, it launches an Alfvén wave that propagates away into the heliosphere. In closed regions, it induces a large-scale twist that is confined to the corona. Third, as shown below, the driving strongly distorts the open-closed boundary, leading to the formation of current sheets and efficient interchange reconnection between open and closed flux. This reconnection has profound consequences for generating the slow wind.

Figure 3 shows the effects of interchange reconnection on the corona and heliosphere. Six field lines traced from fixed footpoints at the solar surface are plotted at three different times. Figure 3A shows these field lines at the end of the driving, $t = 9.5$ hr, when the field lines are closed. Figure 3B shows the field lines traced from the same footpoints only 5 minutes after Figure 3A, and Figure 3C shows them another 5 minutes after that. Also shown are the yellow surface at $1R_\odot$ and the current density from Figure 2, except that now the ecliptic plane is viewed edge-on. The field lines are drawn from footpoints in the northern hemisphere, and the coronal-hole corridor is at Sun center from the viewer's perspective. The box edge is at about $12R_\odot$.

The field-line evolution shown in Figure 3 is the principal result of our simulation. Note that all footpoints at the surface are held fixed during the 10 minutes of the figure; consequently, any changes in the field lines must be due to dynamics within the corona and heliosphere. Additionally, the solar-wind flow speed near the top of the closed-field region is small, $< 100$ km s$^{-1}$. Thus, over the time span of 5 minutes, closed field lines can move less than 30,000 km, which is an imperceptible distance on the scale of the figure. All the changes seen, therefore, must be due to reconnection, not to plasma motion. Even the red field line that remains closed between Figures 3A and 3B must change due to reconnection. Note also the shape of the highest-latitude field lines in Figure 3C. Their curvature is due to the Alfvén wave that was launched in the open flux by the surface motion and has propagated to about $12R_\odot$ by this time. This reemphasizes that the field-line change to open connectivity is due to reconnection, not to outward convection of the field lines from the closed inner corona into the heliosphere.

The striking result of Figure 3 is that, in less than 10 minutes, magnetic flux changes from being closed to open and from being confined near to reaching far from the HCS. Furthermore, the reconnection responsible for these changes occurs high in the corona, near the HCS, at the interface between open and closed flux. The key point is that, at the time of Figure 3C, the coronal sections of the six field lines still contain closed-field plasma, which is now free to flow out along the open heliospheric sections to contribute to the slow wind. Consequently, we conclude that the photospherically driven dynamics of an open-field corridor result in the release of closed-field plasma far from the HCS, as required to explain the observations.

To understand the overall evolution of the corridor magnetic field throughout the post-driving relaxation period, we traced field lines from a grid of footpoints on the surface spanning the coronal-hole corridor and encompassing parts of the closed field to either side. This grid spans [−15°,+15°] longitude and [15°,45°] latitude in the northern hemisphere. Figure 4 shows the location in latitude and longitude on a surface at $12R_\odot$ of the corresponding footpoint of every field line that changed from closed to open. The locations on this surface of each newly opened footpoint are tracked and plotted continuously. This allows us to measure any further reconnection of these field lines. If a field line becomes closed again, the footpoint is no longer shown. To display the time that has elapsed since each footpoint became open, we color the points on the $12R_\odot$ surface by age, where yellow indicates footpoints that have only just become open, and purple indicates footpoints that have been open for nearly a day. Figure 4A shows time $t = 9.6$ hr, immediately after the end of the driving, as the Alfvén wave passes through the surface at $12R_\odot$.

Figures 4B,C,D show times $t = 15.2$ hr, 21.0 hr, and 32.2 hr at the end of the simulation, respectively. An animation of Figure 4 showing the entire simulation is available online (Movie 1).

The first important conclusion from Figure 4 is that recently opened field lines cover almost the complete S-Web arc of Figure 2. There is a gap near the arc apex because the coronal-hole boundary was not deformed across the entire corridor footprint. Consequently, some open flux never experienced interchange reconnection and so is not plotted. For a more realistically complex driver, we expect that all of the flux within the corridor would undergo interchange reconnection and fill the arc. Furthermore, a considerable amount of interchange reconnection occurs during the preceding driving phase and is not captured in Figure 4. Even with that exclusion, it is clear that a substantial amount of flux that previously was closed has opened and reaches high above the HCS.

Another important conclusion is that footpoints continue to open throughout the evolution, even almost a full day after the driving stops. In the yellow locations, along the left side of the arc up to 30° and along the HCS, field lines are constantly opening throughout the simulation. These are locations where closed-field, slow-wind plasma would be released continuously. The purple field lines at the top of the left leg of the arc, above 30° latitude, opened soon after the driving ended and then remained open for the rest of the simulation. The connectivity at all these locations changed much more rapidly than solar-wind plasma could flow out to $12R_\odot$. Nevertheless, these are locations where outflowing closed-field plasma could be observed by spacecraft several hours later.

The right leg of the arc in Figure 4 consists mainly of field lines connecting to photospheric footpoints that changed from closed to open early in the evolution, and then remained open. It is important to note that none of the purple points at high latitudes, above 10°, initially appeared at those latitudes. Those points first became open near the HCS through interchange reconnection, and then shifted to higher latitudes through subsequent open-open reconnections. For this evolution to be due to motion of the plasma, velocities 35 km s$^{-1}$ ≲ $v$ ≲ 500 km s$^{-1}$ would be required. The plasma velocities in the legs of the arc are only on the order of 2 km s$^{-1}$. This result demonstrates again that the connectivity at high latitudes changes via fast open-open reconnection. The rapid opening of field lines along the S-Web arc due to reconnection has essentially the same effect as immediate, direct opening of flux throughout this region. Therefore, Figure 4 shows that even a simple and relatively short-lived distortion of the coronal-hole boundary can release slow-wind plasma all along the arc of Figure 2.

The rapid transition in connectivity along the S-Web arc, which is essential for obtaining closed-field plasma far from the HCS, is the most important but least intuitive aspect of our simulation. Figure 5 uses the mapping of the coronal-hole magnetic field from the surface to the heliosphere to follow this transition, over the same [−15°,+15°]×[15°,45°] domain used to generate Figure 4. Color-coded contours of field-line connectivity are drawn on the solar surface. Black and gray regions are closed: black corresponds to footpoints of field lines that map to the southern hemisphere, gray to footpoints of field lines that map to the northern hemisphere on either side of the coronal-hole corridor. The black field lines in Figure 1C are rooted in the black region immediately adjacent to the boundary of the coronal hole in Figure 5. The white to purple contours correspond to open flux: lighter regions map close to the HCS, while darker regions map to higher latitudes. The two panels in Figure 5 were extracted from the full animation (Movie 2) that is available online.

Figure 5A shows that, before the photospheric driving, the mapping is smooth and well resolved. The S-Web corridor of Figure 1C is clearly distinguished. Its edges, which map to the HCS, are light in color. The gradient is smooth from the edges of the open region to its center, which maps to high latitudes, stepping progressively through higher contour values (also shown by the footpoints and field lines in Figures 1C and 2). For this smooth topology, reconnection between open and closed field lines is possible only for a closed line from the black region, which connects to the south, and an open line from the light-colored region, which maps to the HCS at low latitudes between $0°$ and $5°$. Furthermore, the smoothness of the mapping in the open-flux region indicates that any open-open reconnection in this state would result in a negligible change in field-line latitude. For the topology shown in Figure 5A, therefore, we expect that all opening and, consequently, all slow wind would be constrained to occur very near the HCS.

Figure 5B, in contrast, shows the same contours at the end of the driving but presents a very different scenario. The coronal-hole boundary has been strongly distorted by the photospheric driving, with closed-field regions extending deep into the open-field corridor. The colored contours in the open-field region have also been greatly distorted. Open flux that maps to between

20° and 30° latitude is now directly adjacent to closed flux. Because the mapping between the edges and center of the open region is no longer smooth, interchange reconnection between closed field lines and open lines that map to high latitudes is now possible. Furthermore, the mapping within the open-field region has become quasi-singular, so that open-open reconnection can rapidly change the latitudes of field lines. This quasi-singular mapping of flux within the open-field corridor, after it has been stirred by the photospheric driving, is the underlying reason that closed-field plasma can appear all along the S-Web arc.

On the Sun, the field-line mapping is certain to be even more singular than that shown in Figure 5B, for two reasons. First, convective cells appear and disappear randomly over the whole surface. Second, the Sun's surface flux distribution always is far more complex than the simple, single corridor of Figure 1. Observed photospheric distributions generally produce coronal holes in many locations and at many scales. Such a network of coronal holes creates an equally complex network of slow solar-wind arcs, which are capable of covering a wide band surrounding the HCS (Antiochos et al. 2011; Crooker et al. 2012). Our simulations prove that the dynamics along these S-Web arcs can release slow-wind plasma from the closed-field corona to high latitudes in the heliosphere, arguably to an even greater extent than predicted previously from static models. We conclude that the giant arcs of slow wind implied by Figures 2, 4, and 5 must be a fundamental and pervasive feature of the heliosphere.

**Acknowledgments**

This work was supported by NASA's Living With a Star, Heliophysics Supporting Research, and High-End Computing programs.

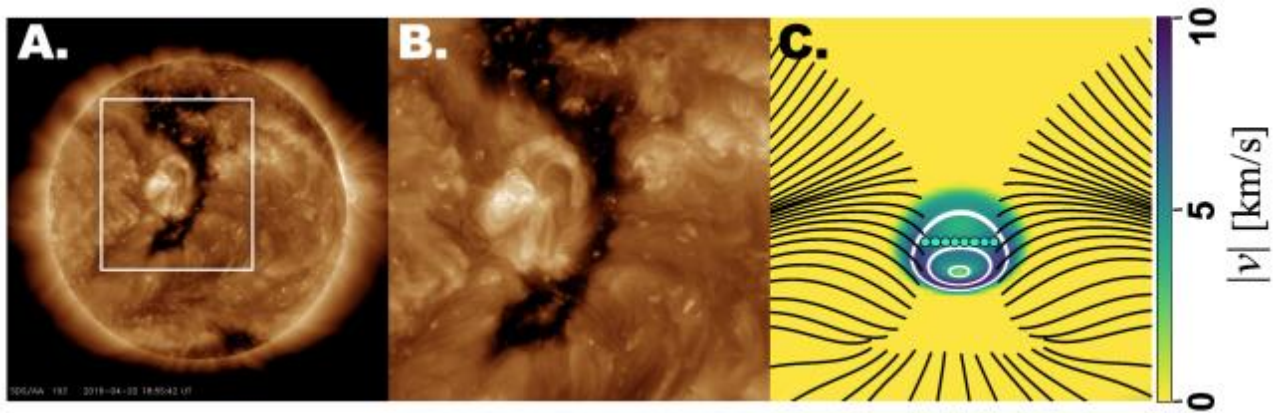

**Figure 1A.** *Solar Dynamics Observatory* EUV image of the Sun. The coronal hole appears dark relative to the surrounding corona. **B.** Zoom-in of white box from A. The coronal hole extends from polar to low latitudes through a narrow open-field corridor at mid latitudes. **C.** Simulated S-Web corridor. Black magnetic field lines outline the boundary of the polar and low-latitude open regions and the connecting open-field corridor on the surface. Cyan dots cross the corridor. Yellow/green/purple shading indicates the driving velocity magnitude applied in the simulation; white velocity streamlines on the surface show the rotational motion.

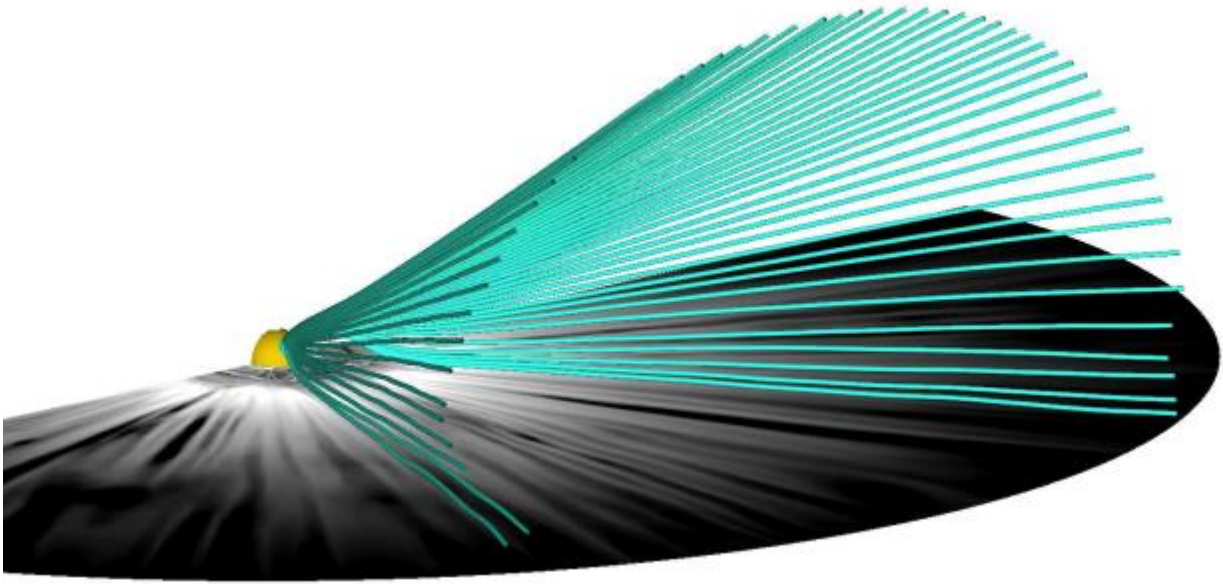

**Figure 2.** Initial simulated S-Web arc. The yellow surface is at $1R_\odot$; cyan magnetic field lines start in the coronal-hole corridor along the line of cyan dots in Figure 1C and end at $30R_\odot$. Gray-scale contours on the ecliptic plane show the current density within the HCS.

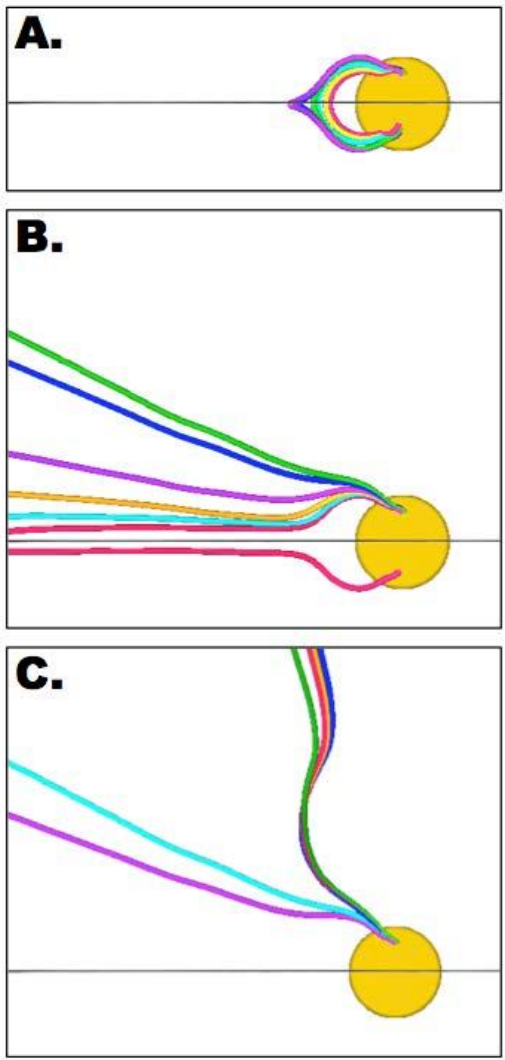

**Figure 3.** Six magnetic field lines traced from fixed footpoints at the edge of the coronal-hole corridor in the northern hemisphere. **A.** Immediately after the driving on the surface has stopped at $t = 9.5$ hr, all field lines are closed. **B.** Five minutes after A, five of the field lines are open. **C.** Five minutes after B, all field lines are open; four reach beyond 30° latitude.

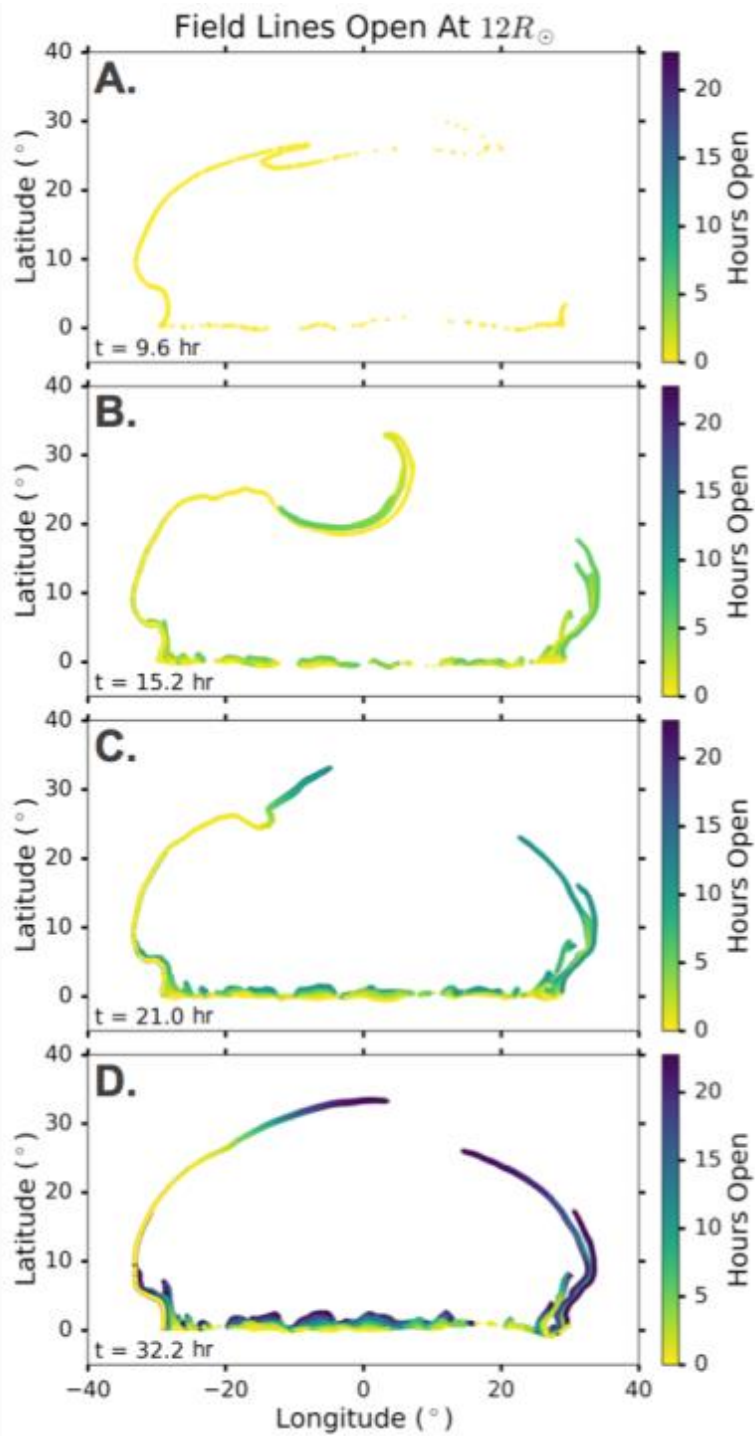

**Figure 4.** Location of field lines on the spherical surface at $12R_\odot$. Only field lines that opened after the driver stops are shown; the color scale represents how long they have been open. Times are:

**A**. $t = 9.6$ hr; **B**. $t = 15.2$ hr; **C**. $t = 21.0$ hr; **D**. $t = 32.2$ hr. The full animation (Movie 1) is available online.

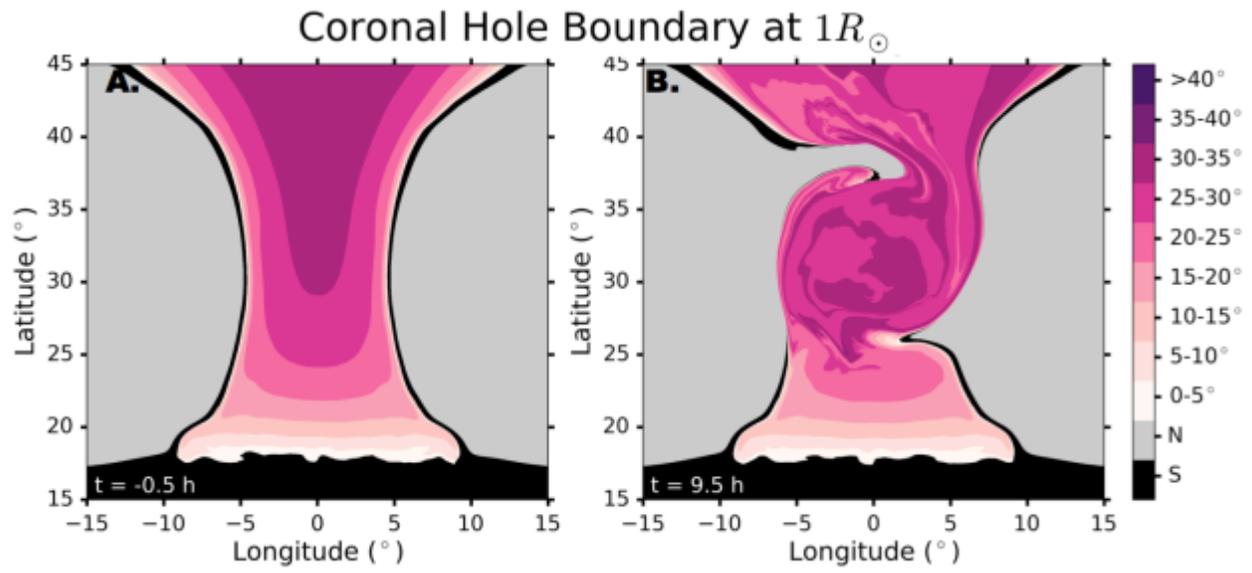

**Figure 5.** Connectivity map of the northern-hemisphere magnetic field on the spherical surface at $1R_\odot$. Field lines rooted in black and gray regions are closed: those in black connect across the equatorial PIL and close in the southern hemisphere; those in gray close across nearby PILs. Field lines rooted in white to purple regions are open, reaching the spherical surface at $12R_\odot$ at latitudes within the indicated bands (0°–5°, etc.). The coronal-hole boundary separates the black and white regions and maps directly to the HCS. Times are: **A.** Before driving, $t = -0.5$ hr; **B.** After driving, at $t = 9.5$ hr. The full animation (Movie 2) is available online.